\documentstyle[preprint,epsfig,aps]{revtex}
\begin{document}

\title{Universal Quantum Computation with the Exchange Interaction}
\author{D. P. DiVincenzo$^1$, D. Bacon$^{2,3}$, J. Kempe$^{2,4,5}$, G.
Burkard$^6$, and K. B. Whaley$^2$}
\address{$^1$IBM Research Division, TJ Watson Research Center, Yorktown
Heights, NY 10598 USA}
\address{$^2$Department of Chemistry, University of California,
Berkeley, CA 94720 USA}
\address{$^3$Department of Physics, University of California, Berkeley,
CA 94720 USA}
\address{$^4$Department of Mathematics, University of California,
Berkeley, CA 94720 USA}
\address{$^5$\'{E}cole Nationale Superieure des
T\'{e}l\'{e}communications, Paris, France}
\address{$^6$Department of Physics and Astronomy, University of Basel,
Klingelbergstrasse 82, CH-4056 Basel, Switzerland}

\maketitle

{\bf Experimental implementations of quantum computer architectures
are now being investigated in many different physical settings.  The
full set of requirements that must be met to make quantum computing a
reality in the laboratory\cite{DPD2000} is daunting, involving
capabilities well beyond the present state of the art.  In this report
we develop a significant simplification of these requirements that can
be applied in many recent solid-state approaches, using quantum dots
\cite{LossBurkard:98}, and using donor-atom nuclear spins
\cite{Kane:98} or electron spins \cite{Vrijen}.  In these approaches,
the basic two-qubit quantum gate is generated by a tunable Heisenberg
interaction (the Hamiltonian is $H_{ij}=J(t){\vec S}_i\cdot{\vec S}_j$
between spins $i$ and $j$), while the one-qubit gates require the
control of a local Zeeman field.  Compared to the Heisenberg
operation, the one-qubit operations are significantly slower and
require substantially greater materials and device complexity, which
may also contribute to increasing the decoherence rate.  Here we
introduce an explicit scheme in which the Heisenberg interaction alone
suffices to exactly implement any quantum computer circuit, at a price
of a factor of three in additional qubits and about a factor of ten in
additional two-qubit operations.  Even at this cost, the ability to
eliminate the complexity of one-qubit operations should accelerate
progress towards these solid-state implementations of quantum
computation.}

The Heisenberg interaction has many attractive
features\cite{LossBurkard:98,LossBurkard:99} that have led to its
being chosen as the fundamental two-qubit interaction in a large
number of recent proposals: Its functional form is very accurate ---
deviations from the isotropic form of the interaction, arising only
from relativistic corrections, can be very small in suitably chosen
systems.  It is a strong interaction, so that it should permit very
fast gate operation, well into the GHz range for several of the
proposals. At the same time, it is very short ranged, arising from the
spatial overlap of electronic wavefunctions, so that it should be
possible to have an on-off ratio of many orders of
magnitude. Unfortunately, the Heisenberg interaction by itself is not
a universal gate\cite{Barenco}, in the sense that it cannot generate
any arbitrary unitary transformation on a collection of spin-1/2
qubits.  So, every proposal has supplemented the Heisenberg
interaction with some other means of applying independent one-qubit
gates (which can be thought of as time-dependent local magnetic
fields).  But the need to add this capability to the device adds
considerably to the complexity of the structures, by putting
unprecedented demands on ``g-factor'' engineering of heterostructure
materials\cite{NATO,Vrijen}, requiring that strong, inhomogeneous
magnetic fields be applied \cite{LossBurkard:98,LossBurkard:99}, or
involving microwave manipulations of the spins that may be slow and
may cause heating of the device \cite{Vrijen}. These added
complexities may well exact a high cost, perhaps degrading the quantum
coherence and clock rate of these devices by orders of magnitude.

The reason that the Heisenberg interaction alone does not give a
universal quantum gate is that it has too much symmetry: it commutes
with the operators ${\hat S}^2$ and ${\hat S}_z$ (for the total spin
angular momentum and its projection on the $z$ axis), and therefore it
can only rotate among states with the same $S,\ S_z$ quantum numbers.
But by defining coded qubit states, ones for which the spin quantum
numbers always remain the same, the Heisenberg interaction alone {\em
is} universal\cite{Bacon:99b,LV,Kempe:00a}, and single-spin operations
and all their attendant difficulties can be avoided.

Recent work has identified the coding required to accomplish this.
Starting with early work that identified techniques for suppressing
phase-loss mechanisms due to coupling with the
environment\cite{DFSpeople5,DFSpeople1,DFSpeople3}, more recent
studies have identified encodings that are completely immune from
general collective decoherence, in which a single environmental degree
of freedom couples in the same way to all the spins in a block.  These
codes are referred to both as decoherence-free subspaces (and their
generalization, the decoherence-free subsystems)
\cite{DFSpeople4,Bacon:99b,Kempe:00a}, and also as noiseless subspaces
and subsystems\cite{DFSpeople2,Knill:00a,LV}.  The noiseless
properties of these codes are not relevant to the present work; but
they have the desired property that they consist of states with
definite angular momentum quantum numbers.

So, in principle, the problem has been solved: the Heisenberg
interaction alone is universal and can be used for quantum
computation.  However, a very practical question still remains: how
great is the price that must be paid in return for eliminating
single-spin operations?  In particular, how many applications of the
Heisenberg interaction are needed to complete some desired quantum
gate operation?  The only guidance provided by the existing
theory\cite{Bacon:99b,LV,Kempe:00a} comes from a theorem of Solovay
and Kitaev\cite{SK1,SK2,SK3}, which states that ``efficient''
approximations exist: given a desired accuracy parameter $\epsilon$,
the number $N$ of exchange operations required goes like $N\approx
K\log^c(1/\epsilon)$, where $c\approx 4$ and $K$ is an unknown
positive constant.  However, this theorem provides very little useful
practical guidance for experiment; it does not show how to obtain the
desired approximating sequence of exchange operations, and, since $K$
is unknown, it gives no clue of whether the number of operations
needed for a practical accuracy parameter is 10 or 10000.  In the
following we remedy these inadequacies by showing that the desired
quantum logic operations can be obtained exactly using sequences of
exchange interactions short enough to be of practical significance for
upcoming experiments.

In the scheme we analyze here, we use the smallest subspace with
definite angular-momentum quantum numbers that can be used to encode a
qubit; this subspace is made up of three spins.  It should be
noted\cite{Kempe:00a} that in principle the overhead in spatial
resources could be made arbitrarily small: asymptotically the rate of
encoding into such noiseless subsystems converges to unity.  The
space of three-spin states with spin quantum numbers $S=1/2$,
$S_z=+1/2$ is two dimensional and will serve to represent our coded
qubit.  A good explicit choice for the basis states of this qubit are
$|0_L\rangle=|S\rangle|\uparrow\rangle$,
$|1_L\rangle=\sqrt{2/3}|T_+\rangle|\downarrow\rangle-\sqrt{1/3}|T_0\rangle
|\uparrow\rangle$.  Here
$|S\rangle=\sqrt{1/2}(|\uparrow\downarrow\rangle-
|\downarrow\uparrow\rangle)$ is the singlet state of spins 1 and 2
(see Fig.  1a) of the three-spin block, and
$|T_+\rangle=|\uparrow\uparrow\rangle$ and
$|T_0\rangle=\sqrt{1/2}(|\uparrow\downarrow\rangle+
|\downarrow\uparrow\rangle)$ are triplet states of these two spins.
For these states we have constructed an explicit exchange
implementation of the basic circuit elements of quantum
logic\cite{Barenco}; in particular, we discuss how one obtains any
coded one-qubit gate, and a specific two-qubit gate, the controlled
NOT (cNOT).

It is easy to understand how one-qubit gates are performed on a single
three-spin block.  We note that Hamiltonian $H_{12}$ generates a
rotation $U_{12}=\exp(i/\hbar\int J{\vec S}_1\cdot{\vec S}_2 dt)$
which is just a $z$-axis rotation (in Bloch-sphere notation) on the
coded qubit, while $H_{23}$ produces a rotation about an axis in the
{\em x-z} plane, at an angle of 120$^o$ from the $z$-axis.  Since
simultaneous application of $H_{12}$ and $H_{23}$ can generate a
rotation around the $x$-axis, three steps of 1D parallel operation
(defined in Fig. 1a) permit any one-qubit rotation, using the classic
Euler-angle construction.  In serial operation, we find numerically
that four steps are always adequate when only nearest-neighbor
interactions are possible (eg, the sequence
$H_{12}$-$H_{23}$-$H_{12}$-$H_{23}$ shown in Fig. 2a, with suitable
interaction strengths), while three steps suffice if interactions can
be turned on between any pair of spins (eg,
$H_{12}$-$H_{23}$-$H_{13}$, see Fig. 2b).

We have performed numerical searches for the implementation of
two-qubit gates using a simple minimization algorithm.  Much of the
difficulty of these searches arises from the fact that while the four
basis states $|0_L,1_L\rangle|0_L,1_L\rangle$ have total spin quantum
numbers $S=1$, $S_z=+1$, the complete space with these quantum numbers
for six spins has nine states, and exchanges involving these spins
perform rotations in this full nine-dimensional space.  So, for a
given sequence, eg the one depicted in Fig.~2c, one considers the
resulting unitary evolution in this nine-dimensional Hilbert space as
a function of the interaction times $t_1$, $t_2$, ... $t_N$.  This
unitary evolution can be expressed as a product
$U(t_1,\dots,t_N)=U_N(t_N)\cdots U_2(t_2)U_1(t_1)$, where
$U_n(t_n)=\exp(it_n H_{i(n),j(n)}/\hbar)$.  The objective of the
algorithm is to find a set of interaction times such that the total
time evolution describes a cNOT gate in the four-dimensional logic
subspace $U(t_1,\dots,t_N)=U_{\rm cNOT}\oplus A_5$.  The matrix $A_5$
can be any unitary $5\times 5$ matrix (consistent with $U$ having a
block diagonal form).  The efficiency of our search is considerably
improved by the use of two invariant functions $m_{1,2}(U)$ identified
by Makhlin \cite{Makhlin:00}, which are the same for any pair of
two-qubit gates that are identical up to one-qubit rotations.  It is
then adequate to use an algorithm that searches for local minima of
the function $f(t_1,\dots,t_N)=\sum_i(m_i(U_{\rm
cNOT})-m_i(U(t_1,\dots,t_N)))^2$ with respect to $t_1$, ...$t_N$
($m_i$ is understood only to act on the $4\times 4$ logic subspace of
$U$).  Finding a minimum for which $f=0$ identifies an implementation
of cNOT (up to additional one-qubit gates, which are easy to
identify\cite{Makhlin:00}) with the given sequence $(i(n),j(n))_n$,
$i(n)\neq j(n)$ of exchange gates.  If no minimum with $f=0$ is found
after many tries with different starting values (ideally mapping out
all local minima), we have strong evidence (although not a
mathematical proof) that the given sequence of exchange gates cannot
generate cNOT.

The optimal serial-operation solution is shown in Fig. 2c.  Note that by
good
fortune this solution happens to involve only nearest neighbors in the
1D
arrangement of Fig. 1a.
The circuit layout shown obviously has a high degree of
symmetry; however, it does not appear possible to give the obtained
solution in a closed form.
(Of course, any gate sequence involving non-nearest neighbors can be
converted
to a local gate sequence by swapping the involved qubits, using the SWAP
gate,
until they are close; here however the {\it minimal} solution found does
not
require such manipulations.)
We have also found (apparently) optimal numerical solutions for parallel
operation mode.  For the 1D layout of Fig. 1a, the simplest solution
found
involves 8 clock cycles
with just 8*4 different interaction-time parameters ($H_{12}$ can
always be zero in this implementation).  For the 2D parallel mode of
Fig. 1b, a solution was found using just 7 clock cycles (7*7
interaction times).

It is worthwhile to give a complete overview of how quantum
computation would proceed in the present scheme.  It should begin by
setting all the computational qubits to the $|0_L\rangle$ state.  This
state is easily obtained using the exchange interaction: if a strong
$H_{12}$ is turned on in each coded block and the temperature made
lower than the strength $J$ of the interaction, these two spins will
equilibrate to their ground state, which is the singlet state.  The
third spin in the block should be in the $|\uparrow\rangle$ state,
which can be achieved by also placing the entire system in a
moderately strong magnetic field $B$, such that $k_B T<<g\mu_B B<J$.
Then, computation can begin, with the one- and two-qubit gates
implemented according to the schemes mentioned above.  For the final
qubit measurement, we note that determining whether the spins 1 and 2
of the block are in a singlet or a triplet suffices to perfectly
distinguish\cite{NATO} $|0_L\rangle$ from $|1_L\rangle$ (again, the
state of the third spin does not enter). Thus, for example, the AC
capacitance scheme for spin measurement proposed by Kane
\cite{Kane:98} is directly applicable to the coded-qubit measurement.

There are several issues raised by this work that deserve further
exploration.  The $S=1/2$, $S_z=+1/2$ three-spin states that we use
are a subspace of a decoherence-free subsystem that has been suggested
for use in quantum computing by exchange interactions
\cite{Kempe:00a,Knill:00a}.  Use of this full subsystem, in which the
coded qubit can be in any mixture of the $S_z=+1/2$ and the
corresponding $S_z=-1/2$ states, would offer immunity from certain
kinds of interactions with the environment, and would not require any
magnetic field to be present, even for initialization of the qubits.
In this modified approach, the implementation of one-qubit gates is
unchanged, but the cNOT implementation must satisfy additional
constraints -- the action of the exchanges on both the $S=1$ and the
$S=0$ six-spin subspaces must be considered.  As a consequence,
implementation of cNOT in serial operation is considerably more
complex; our numerical studies have failed to identify an
implementation (even a good approximate one) for sequences of up to 36
exchanges (cf. 19 in Fig. 2c).  On the other hand, we have found
implementations using 8 clock cycles for 1D and 2D parallel operation
(again for the 1D case $H_{12}$ can be zero), so use of this larger
Hilbert space may well be advantageous in some circumstances.

Finally, we note that further work is needed on the performance of
quantum error correction within this scheme.  Our logical qubits can
be used directly within the error correction codes that have been
shown to produce fault tolerant quantum computation\cite{Preskill}.
Spin decoherence will primarily result in ``leakage'' errors, which
would take our logical qubits into states of different angular
momentum (eg, $S=3/2$).  Our preliminary work indicates that, with
small modifications, the conventional error correction circuits will
not cause uncontrolled propagation of leakage error.  In addition, the
general theory\cite{Lidar:99,Preskill,Bacon:99b,Kempe:00a} shows that
there exist sequences of exchange interactions which directly correct
for leakage by swapping a fresh $|0_L\rangle$ into the coded qubit if
leakage has occurred, and doing nothing otherwise; we have not yet
identified numerically such a sequence.  If fast measurements are
possible, teleportation schemes can also be used in leakage
correction.

To summarize, the present results offer a new alternative route to the
implementation of quantum computation.  The tradeoffs are clear: for
the price of a factor of three more devices, and about a factor of ten
more clock cycles, the need for stringent control of magnetic fields
applied to individual spins is dispensed with.  We are hopeful that
the new flexibility offered by these results will make easier the hard
path to the implementation of quantum computation in the lab.

Acknowledgments: DPD, DB, JK, and KBW acknowledge support from the
National Security Agency (NSA) and the Advanced Research and
Development Activity (ARDA).  DPD also thanks
the UCLA DARPA program on spin-resonance transistors for support, and
is also grateful for the hospitality of D. Loss at the University
of Basel, where much of this work was completed.
JK also acknowledges support from the US National Science Foundation.
The work of GB is supported in part by the Swiss National Science
Foundation.  Discussions with P. O. Boykin and B. M. Terhal are
gratefully acknowledged.

\pagebreak

\vbox{
\begin{figure}
\epsfxsize=14cm
\epsfbox{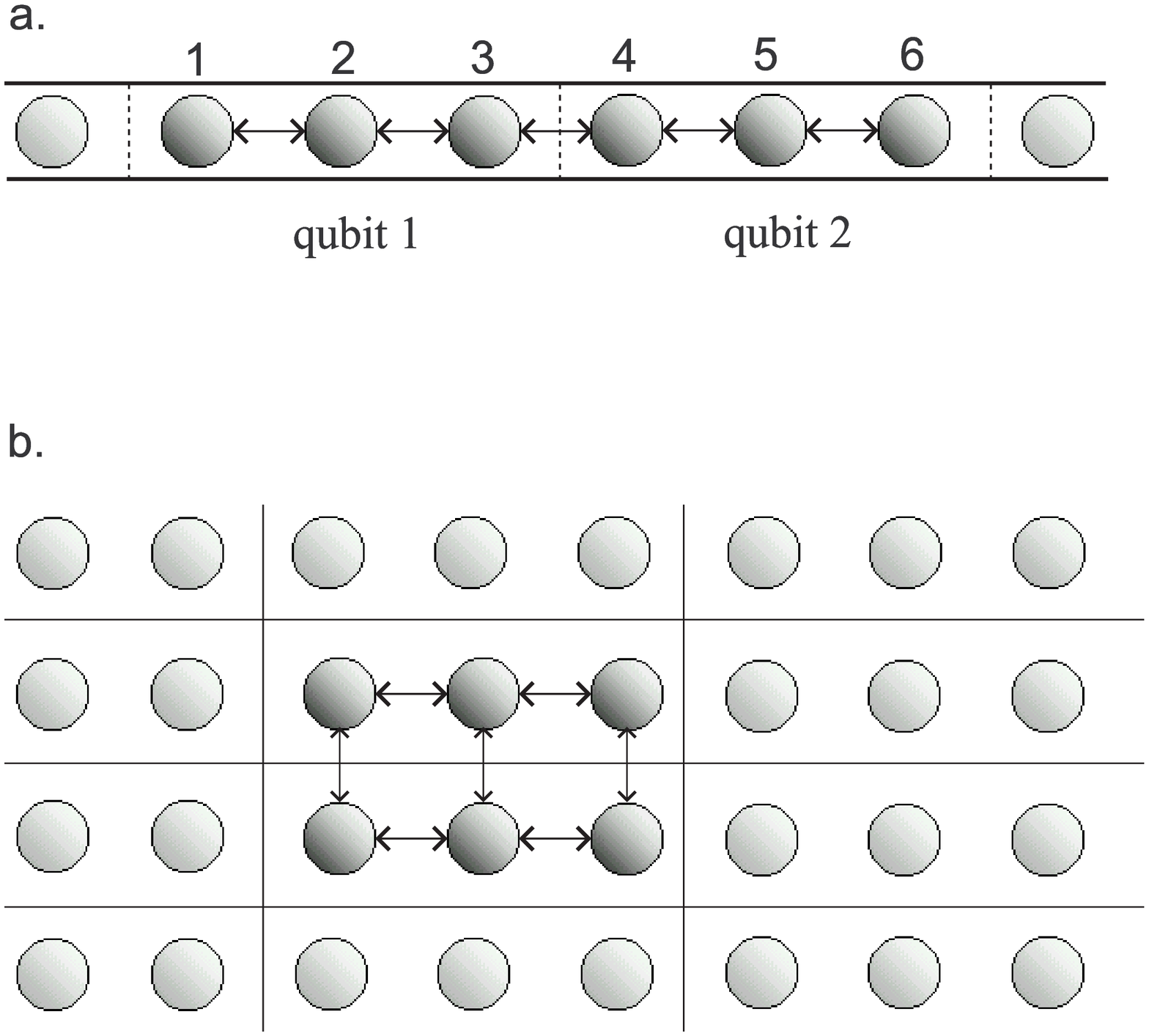}
\smallskip
\caption{Possible layouts of spin-1/2 devices.  a) One-dimensional
layout.  We consider two different assumptions about how the exchange
interactions can be turned on and off in this layout: 1) At any given
time each spin can be exchange-coupled to at most one other spin (we
refer to this as ``serial operation'' in the text), 2) All exchange
interactions can be turned on simultaneously between any neighboring
pair of spins in the line shown (``1D parallel operation'').  b)
Possible two-dimensional layout with interactions in a rectangular
array.  We imagine that any exchange interaction can be turned on
between neighboring spins in this array (``2D parallel operation'').
Of course other arrangements are possible, but these should be
representative of the constraints that will be faced in actual device
layouts.}
\end{figure}
}

\begin{figure}
\epsfxsize=14cm
\epsfbox{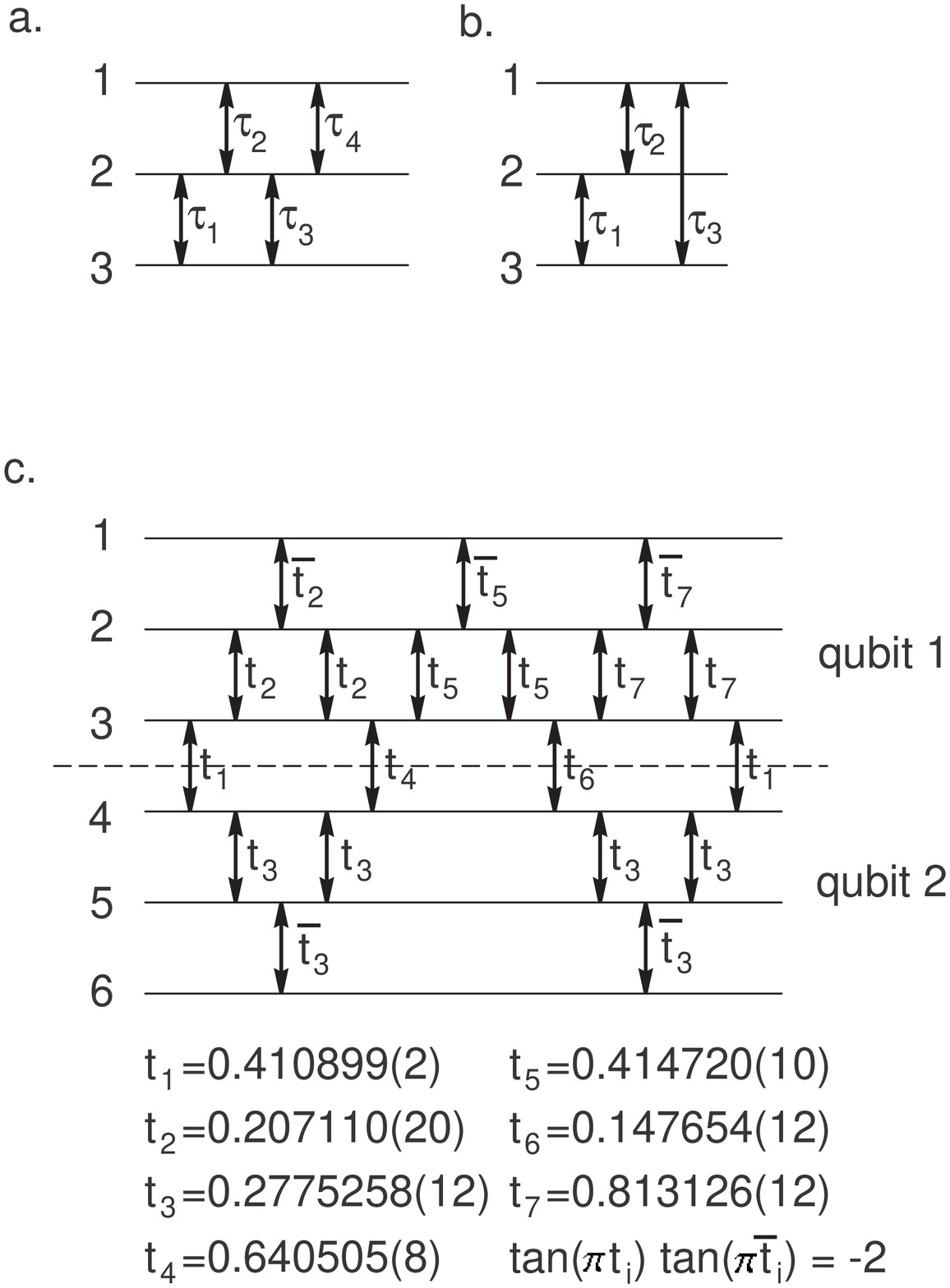}
\smallskip
\caption{Circuits for implementing single-qubit and two-qubit
rotations using serial operations.  a) Single-qubit rotations by
nearest-neighbor interactions.  Four exchanges (double-headed arrows)
with variable time parameters $\tau_i$ are always enough to perform
any such rotation, one of the two possible layouts is shown.  b)
Non-nearest neighbor interactions.  Only three interactions are
needed, one of the possible layouts is shown.  c) Circuit of 19
interactions that produce a cNOT between two coded qubits (up to
one-qubit gates before and after).  The durations of each interaction
are given in units such that for $t=1/2$ the rotation
$U_{ij}=\exp(iJt{\vec S}_i\cdot{\vec S}_j/\hbar)$ is a SWAP,
interchanging the quantum states of the two spins {\em i,j}.  The
$\bar{t}_i$ parameters are not independent, they are related to the
$t_i$s as indicated.  The uncertainty of the final digits of these
times are indicated in parentheses.  With these uncertainties, the
absolute inaccuracy of the matrix elements of the two-qubit gate
rotations achieved is no greater than $6\times 10^{-5}$.  Further fine
tuning of these time parameters would give the cNOT to any desired
accuracy.  In a practical implementation, the exchange couplings
$J(t)$ would be turned on and off smoothly; then the time values given
here provide a specification for the integrated value $\int J(t)dt$.
The functional form of $J(t)$ is irrelevant, but its integral must be
controlled to the precision indicated.  The numerical evidence is very
strong that the solution shown here is essentially unique, so that no
other choices of these times are possible, up to simple permutations
and replacements $t\rightarrow 1-t$ (note that for the Heisenberg
interaction adding any integer to $t$ results in the same rotation).
The results also strongly suggest that this solution is optimal: no
one of these 19 interactions can be removed, and no other circuit
layout with fewer than 19 has been found to give a solution.  We have
also sought, but not found, shorter implementations of other
interesting two-qubit gates like
$\sqrt{\mbox{SWAP}}$\protect\cite{LossBurkard:98,LossBurkard:99}.}
\end{figure}

\end{document}